\documentclass[journal]{IEEEtran}
\usepackage{xcolor}
\usepackage{amsmath}
\usepackage{enumitem}
\usepackage{multirow}
\usepackage{amssymb}
\usepackage{graphicx}
\usepackage{float}
\usepackage{bm}
\usepackage{caption}
\usepackage{subcaption}
\usepackage{hhline}
\usepackage{caption}
\usepackage{tabularx}
\usepackage[ruled,vlined]{algorithm2e}

\usepackage[hidelinks]{hyperref}

\hypersetup{
	colorlinks = True,
	linkcolor=blue,
	citecolor={blue!60!black},
	urlcolor=blue
}
\DeclareMathOperator*{\argmax}{argmax}

\newcommand{\y}{\boldsymbol{y}}
\newcommand{\z}{\boldsymbol{z}}
\newcommand{\X}{\boldsymbol{X}}
\newcommand{\Y}{\boldsymbol{Y}}
\newcommand{\A}{\boldsymbol{A}}
\newcommand{\btheta}{\boldsymbol{\theta}}
\newcommand{\orcid}[1]{\href{https://orcid.org/#1}{\includegraphics[width=8pt]{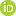}}}
\ifCLASSINFOpdf
\else
\fi
\begin{document}
%
\title{Self-supervised Representation Learning With
\\  Path Integral Clustering For Speaker Diarization}
%
%
%

\author{$\text{Prachi Singh}^\text{\orcid{0000-0003-0760-8572}}$,~\IEEEmembership{Student Member,~IEEE,} and 
        $\text{Sriram~Ganapathy}^\text{\orcid{0000-0002-5779-9066}}$,~\IEEEmembership{Senior Member,~IEEE} 
\thanks{This work was supported   by the grants from  the British Telecom Research Center (BTRC) and with grants from the Department of Atomic Energy project 34/20/12/2018-BRNS/3408.}
\thanks{P. Singh and S. Ganapathy are with the Learning and Extraction of Acoustic Patterns (LEAP) lab, Department of Electrical Engineering, Indian Institute of Science, Bangalore, India, 560012. 
 e-mail: \{prachis, sriramg\}@iisc.ac.in}
 }

%
%

\markboth{Journal of \LaTeX\ Class Files,~Vol.~xx, No.xx, XX , XXXX}%
{Singh \MakeLowercase{\textit{et al.}}: Self-supervised Representation Learning with Path Integral Clustering For Speaker Diarization}
%



\maketitle

\begin{abstract}
Automatic speaker diarization techniques typically involve a two-stage processing approach where audio segments of fixed duration are converted to vector representations in the first stage. This is followed by an unsupervised clustering of the representations in the second stage. In most of the prior approaches, these two stages are performed in an isolated manner with independent optimization steps. In this paper, we propose a representation learning and clustering algorithm that can be iteratively performed for improved speaker diarization. The representation learning is based on principles of self-supervised learning while the clustering algorithm is a graph structural method based on path integral clustering (PIC). The representation learning step uses the cluster targets from PIC and the clustering step is performed on embeddings learned from the self-supervised deep model. This iterative approach is referred to as self-supervised clustering (SSC).  The diarization experiments are performed on CALLHOME and AMI meeting datasets. In these experiments, we show that the SSC algorithm improves significantly over the baseline system (relative improvements of $13$\% and $59$\% on CALLHOME and AMI datasets respectively in terms of diarization error rate (DER)). In addition, the DER results reported in this work improve over several other recent approaches for speaker diarization.  \end{abstract}


%
\IEEEpeerreviewmaketitle

\section{Introduction}

%
\IEEEPARstart{S}{peaker} diarization, the task of partitioning the input audio stream into segments based on speaker sources, has gained substantial interest in the recent years.  The challenges in speaker diarization arise from short speaker turns, background noise, overlapping speech etc. Some of these challenges are highlighted in the recent DIHARD evaluations \cite{ryant2018first,ryant2019second,ryant2020third}.  The early approaches to speaker diarization involved steps of change detection and gender classification~\cite{tranter2006overview}. The recent efforts   have been directed towards a two stage approach of deriving embeddings from  short segments followed by a clustering of the embeddings. The i-vector embedding extraction showed promising improvements for speaker diarization \cite{sell2014speaker}. With the advancements in deep learning, the i-vector embeddings were successfully replaced with deep neural network (DNN) based embeddings like the x-vectors \cite{snyder2018x}, which deploy a  time delay neural network (TDNN). The x-vectors are also effective on short recordings~\cite{kanagasundaram2019study}. 

In terms of clustering, the bottom-up clustering has been the most promising for speaker diarization \cite{anguera2012speaker}. The approach aims at successively merging until a stopping criterion is met~\cite{anguera2006robust}.  The agglomerative hierarchical clustering (AHC) \cite{day1984efficient}, a bottom-up approach, is the most popular method used for clustering.  { The mean-shift algorithm \cite{senoussaoui2014} is another approach which iteratively clusters by assigning data points to the means in a local vicinity.} There has been efforts on pre-processing methods applied on embeddings like length normalization \cite{garcia2011analysis}, recording level principal component analysis (PCA) \cite{zhu2016online} as well as the use of probabilistic linear discriminant analysis (PLDA) based affinity matrix computation for AHC ~\cite{sell2014speaker,ramoji2020nplda}. {  Vi$\tilde{\text{n}}$als et al. \cite{Vials2017DomainAO} introduced session specific adaptation of PLDA using unsupervised clustering labels to re-estimate model parameters and speaker labels}. The other approaches for clustering include spectral clustering approach~\cite{ning2006spectral} and the   information bottleneck based approach \cite{vijayasenan2009information}. Some of the recent approaches to speaker diarization have explored clustering free methods~\cite{zhang2019fully} or end-to-end models  with two speaker conversations \cite{fujita2019end}. 
\begin{figure*}[t!]
	\centering
	\includegraphics[trim={0cm 0cm 0cm 0cm},clip,width=0.6\textwidth]{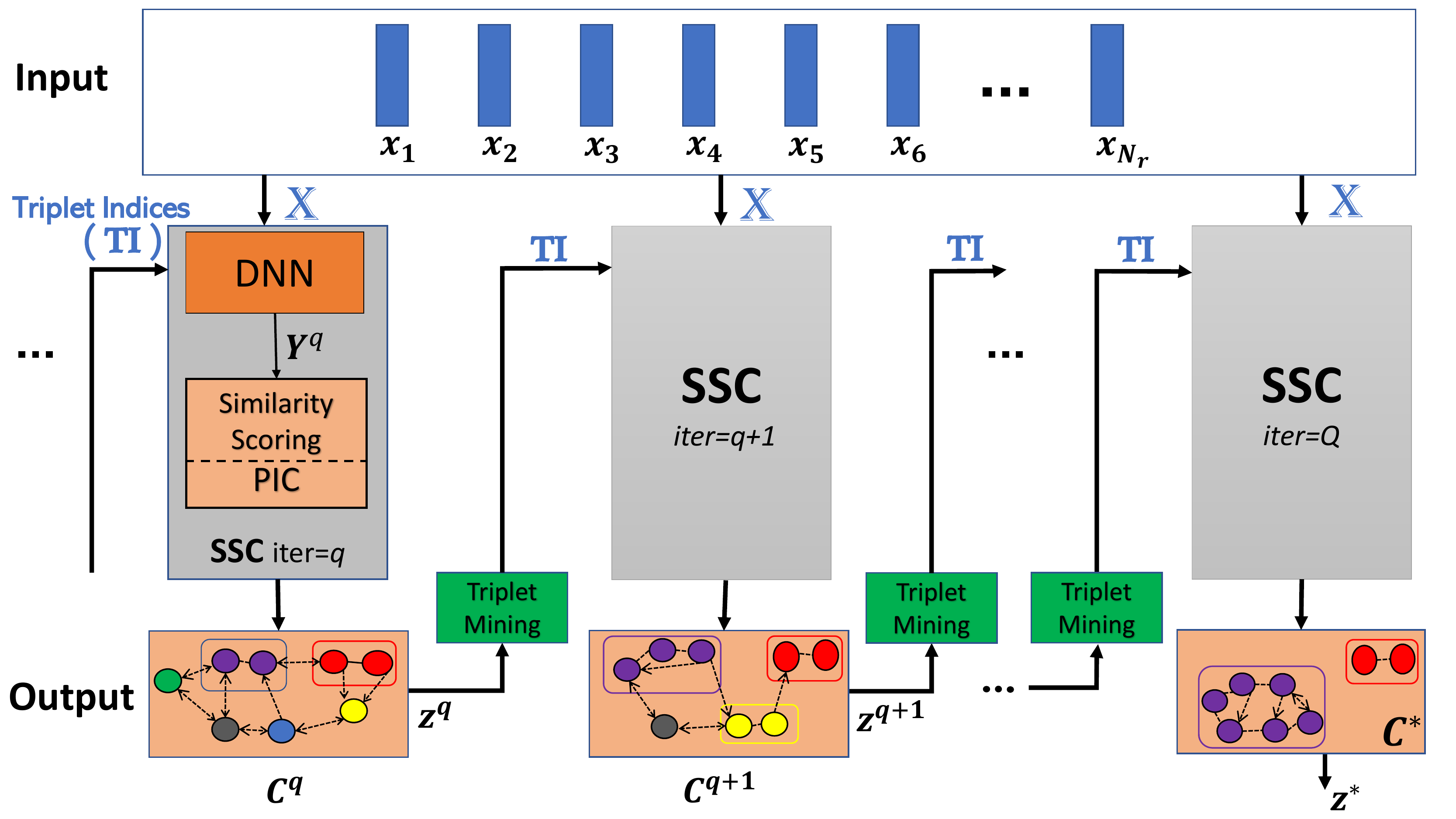}
	\caption{Block schematic of the self-supervised diarization (SSC). The DNN embeddings are used in the path integral clustering (PIC). The clustering outputs from PIC create labels $\bm{z}^q$ used in sampling triplets for iteration $(q+1)$.  The bottom row - circles represent embeddings, colors represent clusters and the dashed lines correspond to graph edges in PIC.}
	\label{fig:schematic}
\end{figure*}
 
In this paper, we propose an approach based on self-supervised representation learning for speaker diarization. Self-supervised learning is a branch of unsupervised learning where the data itself provides supervision labels for model learning ~\cite{hendrycks2019using}. Self-supervised learning has been attempted for phoneme and speech recognition tasks~\cite{pascual2019learning}.
 For speaker diarization, the proposed approach involves iteratively updating embedding representations and cluster identities. The approach alternates between merging the clusters for a fixed embedding representation and learning the representations using the given cluster labels. We show that the proposed approach can  be applied to traditional AHC and to graph based path integral clustering (PIC) \cite{zhang2013agglomerative}. 
 
This paper extends the previous work \cite{Singh2020} using path integral clustering (PIC) and for the tasks with unknown number of speakers. We refer to the proposed approach of joint representation learning and clustering as self-supervised clustering (SSC). The key contributions from this work are: 
 \begin{itemize}
    \item A simple method to jointly learn deep representations and speaker clusters from an unlabeled recording.
    \item Using graph based PIC for  computing affinity measures between clusters.
    \item A stopping criterion based on eigen-values of the affinity matrix.
    \item Incorporating a temporal continuity criterion in the SSC to improve the smoothness of the clustering decisions.
\end{itemize}    
The representation learning framework in the proposed SSC approach uses a triplet similarity based learning. The diarization experiments are performed on the CALLHOME (CH)~\cite{callhome} and Augmented Multi-party Interactions (AMI)~\cite{mccowan2005ami} datasets. Using the proposed SSC approach, we illustrate significant improvements over the x-vector AHC baseline system. The proposed approach also shows advancements over other published results on these datasets. Furthermore, the proposed approach can be used as an initialization for frame-level refinement based on variational Bayes (VB) hidden Markov model (HMM) \cite{diez2018speaker, singh2019leap}. \\



The rest of the paper is organized as follows. Section~\ref{sec:relatedWork} gives an account of related prior work. Section~\ref{sec:proposedWork} details the proposed SSC clustering. The dataset details are given in Section~\ref{sec:data}. The experiments on CALLHOME (CH) dataset and AMI dataset are reported in Section~\ref{sec:CH} and Section~\ref{sec:AMI} respectively. This is followed by an analysis in Section~\ref{sec:analysis}. The paper concludes with a summary  in Section~\ref{sec:concl}.

\section{Related Work}\label{sec:relatedWork}
Early works on diarization used short-term spectral feature vectors and unsupervised clustering based on Gaussian mixture models (GMMs)~\cite{reynolds2005approaches} or hidden Markov models (HMMs)~\cite{wooters2007icsi}. In the last decade, embeddings  based on dimensionality reduction of GMM statistics, termed as i-vectors~\cite{dehak2010front}, have been explored for speaker diarization tasks~\cite{sell2014speaker}. The  i-vector embeddings are extracted from relatively long ($1$-$2$ sec.) segments of audio and they are clustered using   Agglomerative Hierarchical Clustering (AHC)~\cite{day1984efficient}.
 The affinity measure used initially was the cosine similarity score~\cite{silovsky2012speaker}. The model based affinity scoring using probabilistic linear discriminant analysis (PLDA) \cite{ioffe2006probabilistic}, which was successful in speaker verification \cite{lei2012towards}, has also been incorporated for speaker diarization with AHC~\cite{sell2014speaker}. In the recent years, x-vector embeddings~\cite{snyder2018x} trained using time delay neural networks (TDNNs) have replaced i-vectors for PLDA based AHC~\cite{garcia2017speaker}.  The use of recurrent neural networks (RNN) has also been explored for embedding extraction~\cite{wang2018speaker,cyrta2017speaker}.

In terms of the back-end modeling for speaker diarization, in addition to AHC, k-means clustering  \cite{shum2011exploiting}, and spectral clustering have also been explored~\cite{wang2018speaker}.  The LSTM affinity measure has also been proposed for speaker diarization~\cite{Lin2019}. A fully supervised speaker diarization system has been investigated using unbounded interleaved RNN by Zhang et. al.~\cite{zhang2019fully}. The model is trained on conversational data directly with speaker labels. Separately, the application of end-to-end modeling for two speaker conversational data has been explored in \cite{fujita2019end}. In the end-to-end learning, the input features are fed to a model where the loss is either permutation-invariant cross entropy \cite{horiguchi2020end, li2019discriminative} or clustering based \cite{pal2020speaker}. Further to refine the boundaries of segmentation output in speaker diarization, a second re-segmentation step involving frame-level ($20$-$30$ms) modeling ~\cite{diez2018speaker,singh2019leap} can be performed. This model the variational-Bayes HMM model which is typically initialized using a segment level diarization output~\cite{diez2019analysis}. 

The loss functions in clustering of images and text documents have explored cluster assignment losses such as k-means loss \cite{yang2017towards}, spectral clustering loss~\cite{shaham2018spectralnet} and agglomerative clustering loss~\cite{yang2016joint}. These losses are derived from the unlabeled data itself and hence are self-supervised. 

In this paper, we propose a novel approach for  representation learning and clustering based on self-supervised learning. This approach, termed self-supervised clustering (SSC), is inspired by  Yang et. al~\cite{yang2016joint}, which explores  representation learning using agglomerative clustering based loss.
The clustering is a forward operation while the representation learning is a backward operation. The motivation for this framework comes from the intuition that succinct speaker representations are beneficial for clustering while clustering results can provide self-supervisory targets for representation learning.  The agglomerative clustering in the proposed SSC approach is based on path integral clustering (PIC), a graph-structural algorithm \cite{zhang2013agglomerative}.

\section{Self-supervised Clustering}\label{sec:proposedWork}
The block schematic of the self-supervised clustering (SSC) algorithm is given in Figure \ref{fig:schematic}. 
The inputs to the model are x-vector embeddings extracted from fixed length audio segments (details of the x-vector embedding extraction are given in Section~\ref{sec:data}). The deep neural network (DNN) in Figure~\ref{fig:schematic} is a two layer network. The output layer of the DNN generates representations that are used in the clustering. 

The SSC implements iterative steps of clustering and representation learning. The clustering (forward operation) is performed using path integral clustering (described in Section~\ref{sec:pic}). The representation learning (backward operation) is performed using the DNN training with modified triplet similarity (described in Section~\ref{sec:triplet_loss}). 
The iterative process is repeated till the required number of clusters is reached or the stopping criterion is met (discussed in Section~\ref{sec:stopping_criterion}). 
\subsection{Notation}
The model parameters of the representation learning deep neural network (DNN) are denoted by $\boldsymbol{\theta}$. Let $q$ denote the iteration index.  Further, let \begin{itemize}[leftmargin=*]
    \item $\textbf{X}_r = \{\bm{x}_{1},...,\bm{x}_{N_r}\} \in \mathbb{R}^{D}$ denote the sequence of $N_r$ input x-vector features for the recording $r$.  
    \item $\mathbf{z}^q=\{z_1^q,...,z_{N_r}^q\}$ denote the cluster labels for this recording,  where each element takes a discrete cluster index value. \item $\textbf{Y}^q=\{\bm{y}_1^q,...,\bm{y}_{N_r}^q\} \in \mathbb{R}^{d}$ denote the output representations from the DNN model. 
    \item $\textbf{C}^q=\{\mathcal{C}_1^q,...,\mathcal{C}_{N^q}^q\}$ denote the cluster set where each cluster $\mathcal{C}_i^q=\{\bm{y}_k^q|z_k^q=i,\forall k\in \{1,..,N_r\} \}$. 
\item $\mathcal{A}$ denote the affinity  measure  between two clusters. 
\item $\textbf{S}\in\mathbb{R}^{N_rXN_r}$ denote pairwise similarity score matrix obtained using pair-wise similarity $s(i,j)$ between representations at time index $i$ and $j$.
\item $N^q$ denote the number of clusters at $q$-th iteration
\item $N^*$ denote the target number of clusters in the algorithm.
\item $Q$ denote total number of iterations in the algorithm.
\item Hyper-parameters: \\
$K$ - Number of nearest-neighbors used in PIC (Equation~(\ref{eq:trans_prob}))\\
$\sigma$ - scaling factor in the computation of path integral in PIC (Equation~(\ref{eq:path_integral}))\\
$\alpha$ - weighting factor in DNN learning (Equation~(\ref{eq:triplet_loss}))\\
$\phi^q$ - threshold to estimate $N^q$.\\
$\beta$, $n_b$ - positive decaying factor and floor value respectively (Equation~(\ref{eq:temp_continuity}))

\end{itemize}
\subsection{Background - Agglomerative Clustering} \label{sec:AHC}
The AHC operates directly on short-segment embeddings (for the baseline system, these are the  input x-vectors $\X_r$) and the algorithm merges two clusters at each step. The clusters to be merged at the $q$-th iteration are determined using, 
\begin{equation}
\left\{\mathcal{C}_{a}, \mathcal{C}_{b}\right\}=\underset{C_{i}^q, C_{j}^q \in \mathcal{\textbf{C}}^q, i \neq j}{\operatorname{argmax}} \mathcal{A}\left(\mathcal{C}_{i}^q, \mathcal{C}_{j}^q\right)
\label{eq:ahc_cluster_affinity}
\end{equation}
The selected clusters $\left\{\mathcal{C}_{a}, \mathcal{C}_{b}\right\}$ are merged to form a new cluster $\left\{\mathcal{C}_{a} \cup  \mathcal{C}_{b}\right\}^{q+1}$  for the next iteration. The merging process is stopped when the  stopping criteria is met. The affinity $\mathcal{A}$ between clusters can be defined in multiple ways. It is computed using pairwise similarity between embeddings present in the similarity score matrix $\textbf{S}$. The similarity score ($s(i,j)$) in  state-of-the-art diarization systems use the PLDA modeling~\cite{garcia2017speaker}, while in the proposed SSC algorithm we use the cosine similarity measure.  In conventional AHC, the affinity between two clusters can be defined using different linkage choices~\cite{Hastie09-TEO}.\\ $\mathcal{A}\left(\mathcal{C}_{a}, \mathcal{C}_{b}\right)$ can be defined using:
\begin{itemize}
    \item Single-linkage (Nearest-neighbor): The highest similarity score between two elements (one element from each cluster). $max_{i,j}\{s(i,j): \bm{x}_i \in \mathcal{C}_a, \bm{x}_j \in \mathcal{C}_b\}$
     \item Complete-linkage (Farthest-neighbor): The lowest similarity score between two elements (one element from each cluster).  $min_{i,j}\{s(i,j): \bm{x}_i \in \mathcal{C}_a, \bm{x}_j \in \mathcal{C}_b\}$
     \item Average-linkage : The average similarity score of all pairs of elements (one from each cluster).  $mean_{i,j}\{s(i,j): \bm{x}_i \in \mathcal{C}_a, \bm{x}_j \in \mathcal{C}_b\}$
\end{itemize}
In our baseline experiments with AHC, we have used average linkage as the cluster affinity measure in Equation~(\ref{eq:ahc_cluster_affinity}). The use of pairwise similarities for affinity may fail to capture the global structure of data~\cite{pavan2006dominant}. This can be addressed in the graph based agglomerative clustering  algorithms. 
\begin{figure*}[t!]
	\centering
	\includegraphics[trim={0cm 8cm 0cm 2cm},clip,width=0.8\linewidth]{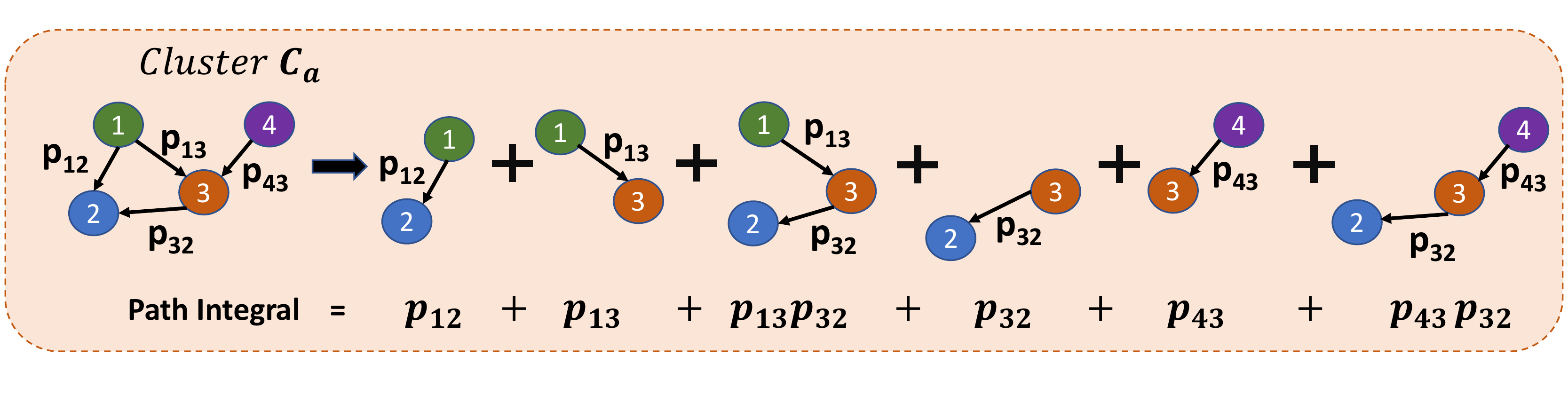}
	\caption{Computation of path integral of a graph. Cluster $C_a$ is represented as graph with edge weights as transition probabilities obtained using Equation~(\ref{eq:trans_prob}). Path integral is the sum of probabilities of all possible paths in the graph. }
	\label{fig:pic}
	\vspace{-10pt}
\end{figure*}
\begin{figure}[t!]
	\centering
	\includegraphics[trim={5cm 0cm 5cm 2cm},clip,width=\linewidth]{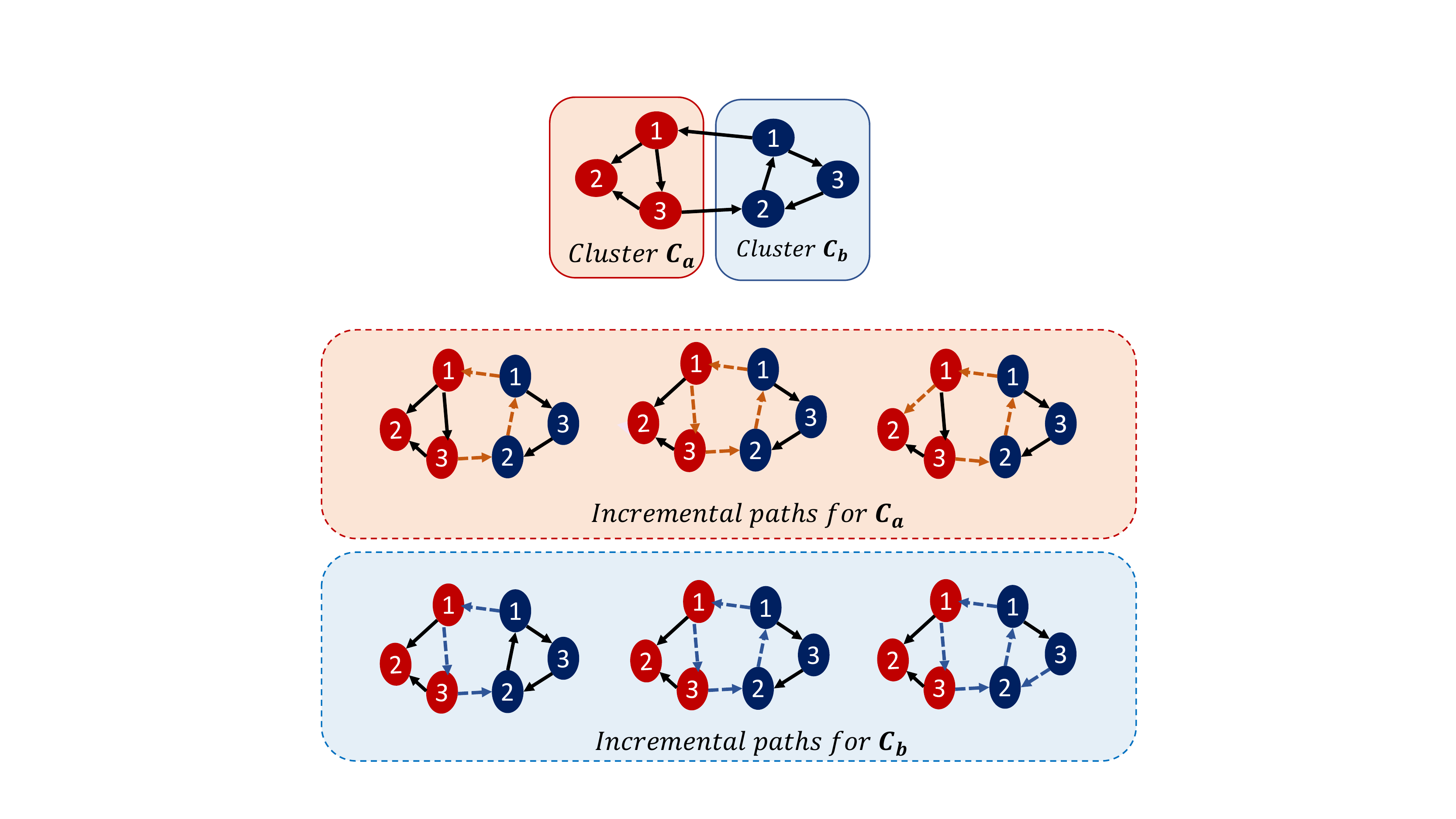}
	\vspace{-2pt}
	\caption{Illustration of incremental paths of clusters $C_a$ and $C_b$ for computing affinity using Equation~(\ref{eq:affnty_pic}). Second and third rows show paths for $C_a \cup C_b$. Second row highlights paths which start and end in cluster $C_a$  with purple dashed arrows. Third row shows the paths which start and end in $C_b$ with blue dashed arrows.}
	\label{fig:inc_pic}
\end{figure}
\subsection{Path Integral Clustering} \label{sec:pic}
Path integral clustering (PIC)~\cite{zhang2013agglomerative} is a graph-structural agglomerative clustering algorithm~\cite{pavan2006dominant} where the graph encodes the structure of the embedding space. 
It exploits the connectivity of directed graph to efficiently cluster the embeddings. The PIC has been attempted for image segmentation and document clustering~\cite{tang2015saliency}. 

The PIC uses path integral as a  structural descriptor of clusters. The clustering process is treated as a dynamical system, with vertices as input features (states of the system) and edge weights as transition probabilities between states. The affinity between two clusters is defined as the incremental path integral when the two clusters are merged. This affinity is based on the assumption that, if two clusters are closely connected, the stability of system can be enhanced by merging the two clusters. 

The PIC algorithm consists of the following steps, 
\begin{itemize}[leftmargin=*,listparindent=\parindent]
    \item Defining the digraph: We build the directed graph based on similarity scores between embeddings. The digraph is denoted as  $G = (V,E)$, where  $V$ is the set of vertices corresponding to the embeddings in $\Y$ and $E$ is the set of edges connecting the vertices. The adjacency matrix  $\bm{W}\in\mathbb{R}^{N_rXN_r}$ is a sparse matrix defined as,   
\begin{equation}
    [W]_{ij}=\begin{cases}
    { \frac{1}{1+exp(-s(i,j))}}, & \text{if $\bm{y}_j \epsilon N_i^K$}.\\
    0, & \text{otherwise}.
  \end{cases}
  \label{eq:trans_prob}
\end{equation}
 where, $N_i^K$ is the set of $K$-nearest neighbors of $\bm{y}_i$. The transition probability matrix $\boldsymbol{P}$ is obtained from the adjacency matrix $\bm{W}$ by normalizing each row with its sum. The transition probability matrix $\boldsymbol{P}$ contains   the edge weights of the graph. 
 The directed edge connecting node $i$ to node $j$ exists only if the value of $[W]_{ij}$ is non-zero.
\item Cluster initialization: We group the embeddings $\Y$ with their first nearest neighbour to form $N_r/2$ clusters. If the clusters have common elements, these are further merged.

\item Defining path integral: We define path integral $\mathcal{S}_{\mathcal{C}_a}$ of a cluster $\mathcal{C}_a$ as the weighted sum of probabilities of all possible paths from any vertex $i$ to any other vertex $j$, where $i,j$ vertices belong to the cluster $\mathcal{C}_a$ and all the vertices along the path also belong to cluster $\mathcal{C}_a$. An illustration of path integral is shown in Figure \ref{fig:pic}. The unnormalized pairwise path integral\footnote{{The proof of the path integral equation is given in Appendix-A}} $g_{i j}$ over all paths (of lengths $k$ ranging from $1$ to $\infty$) from vertex $i$ to $j$ for $\mathcal{G}_{\mathcal{C}_a}$ is given as:\\
\begin{equation}
g_{i j}=\delta_{i j}+\sum_{k=1}^{\infty} \sigma^{k} \sum_{\gamma \in \Gamma_{i j}^{(k)}} \prod_{s=1}^{k} p_{u_{s-1}, u_{s}}
\label{eq:path_integral}
\end{equation}

where, $u_0=i$, $u_k=j$, $p_{u_{s-1}, u_{s}}$ is transition probability from vertex $u_{s-1}$ to $u_{s}$ and  $\delta_{ij}$ is Kronecker delta function defined as $\delta_{ij}=1$ if $i=j$ and $0$ otherwise. $\Gamma_{ij}^{(k)}$ is the set of all possible paths from $i$ to $j$ of length $k$. The constant $0<\sigma<1$ gives more weight to shorter paths compared to longer paths. The path integral $\mathcal{S}_{\mathcal{C}_{a}}$ of the cluster $\mathcal{C}_{a}$ is then the summation of $g_{ij}$ over all pairs of  vertices $i,j \in \mathcal{C}_{a}$ normalized by $|\mathcal{C}_{a}|^2$.  

We define the conditional path integral  $\mathcal{S}_{\mathcal{C}_{a} \mid \mathcal{C}_{a} \cup \mathcal{C}_{b}}$ as the path integral of all paths in $\mathcal{C}_{a} \cup \mathcal{C}_{b}$ such that the paths start and end with vertices belonging to  $\mathcal{C}_{a}$. 
As shown in Appendix \ref{appendix:proof},  the normalized path integral and the normalized conditional path integral converges to  
\begin{eqnarray}
\mathcal{S}_{\mathcal{C}_a} &=& \frac{1}{|\mathcal{C}_a|^2}\bm{1}^T\left(\bm{I}-\sigma\bm{P}_{\mathcal{C}_a}\right)^{-1}\bm{1} \label{eq:pic_ca}\\
\mathcal{S}_{\mathcal{C}_{a} \mid \mathcal{C}_{a} \cup \mathcal{C}_{b}} &=& \frac{1}{|\mathcal{C}_a|^2}\bm{1}_{\mathcal{C}_a}^T\left(\boldsymbol{I}-\sigma\bm{P}_{\mathcal{C}_{a} \cup \mathcal{C}_{b}}\right)^{-1}\bm{1}_{\mathcal{C}_{a}} \label{eq:pic_ca_cb}
\end{eqnarray}
where, $\bm{P}_{\mathcal{C}_a}$ and $\bm{P}_{\mathcal{C}_{a} \cup \mathcal{C}_{b}}$ are the sub-matrices of the transition probability matrix $\bm{P}$ obtained by selecting vertices that belong to $\mathcal{C}_a$ and $\mathcal{C}_{a} \cup \mathcal{C}_{b}$ respectively. Here, $|\mathcal{C}_a|$ denotes the cardinality (\# of vertices) of cluster $\mathcal{C}_a$, $\bm{1}$ is a column vector of all ones of size $|\mathcal{C}_a|$ and $\bm{1}_{\mathcal{C}_a}$ is a binary column vector of size $|\mathcal{C}_a \cup \mathcal{C}_b|$ with ones at all locations corresponding to the vertices of $\mathcal{C}_a$ and zeros at all locations corresponding to the vertices of $\mathcal{C}_b$.  Note that the identity matrix $\bm{I}$ used in Equation~(\ref{eq:pic_ca}) and (\ref{eq:pic_ca_cb}) are of dimensions $|\mathcal{C}_a|$ and $|\mathcal{C}_a \cup \mathcal{C}_b|$ respectively.

\item Cluster merging: The cluster affinity measure for the PIC algorithm is computed as,
\begin{equation}\label{eq:affnty_pic}
\mathcal{A}\left(\mathcal{C}_a,\mathcal{C}_b\right)= [\mathcal{S}_{\mathcal{C}_{a} \mid \mathcal{C}_{a} \cup \mathcal{C}_{b}}-\mathcal{S}_{\mathcal{C}_{a}}]  + [\mathcal{S}_{\mathcal{C}_{b} \mid \mathcal{C}_{a} \cup \mathcal{C}_{b}}-\mathcal{S}_{\mathcal{C}_{b}}] 
\end{equation}
 where the term  inside each square bracket is called incremental path integral. The computation of the incremental path integrals is illustrated in Figure \ref{fig:inc_pic}. A higher value of affinity between two clusters indicates the presence of more dense connections between them. Thus,  merging the clusters with maximum affinity will form a more coherent  cluster. Using the definition of affinity (Equation~(\ref{eq:affnty_pic})),  the clusters with maximum affinity are merged at each iteration as shown in Equation~(\ref{eq:ahc_cluster_affinity}). 
\end{itemize}

\begin{algorithm}[t!]
\SetAlgoLined
\textbf{Initialize:}$(q=0)$\\
$\btheta^0\leftarrow$ (Whitening+PCA)\\
$\Y^0\leftarrow$ PCA outputs\\
If unknown $N^*\rightarrow N^*=1$\\
 \While{continue}{
  \begin{enumerate}
      \item $q=q+1$
      \item Sample triplets based on $\z^{q-1}$
      \item $\bm{\theta}^{q-1}\xrightarrow{\text{DNN with triplet training}}\btheta^q$
      \item $\X_r\xrightarrow{\text{Forward Pass}(\btheta^q)}\Y^q$
      \item $\tilde{N}^q=$Estimate$\_N^q(\Y^q,\phi^q,N^{q-1})$
      \item $N^q=max(N^*,\tilde{N}^q)$
      \item \If{$N^q==N^*$ or $q==Q$}{
      \begin{itemize}
      \item[] $Q=q$
      \item[] $N^*=N^Q$
      \item[] break
      \end{itemize}
      }
      \item $\Y^q\xrightarrow{\text{PIC($N^q$ clusters)}}\z^q=\{z_1^q,...,z_{N_r}^q\}$
  \end{enumerate}
 }
 \textbf{Termination:}\vspace{0.5em} $\X_r\xrightarrow{\text{DNN training + Forward Pass($\btheta^*$)}}\Y^*=\{\y_1^*,...,\y_{N_r}^*\}$\\
 $\{\y_1^*,...,\y_{N_r}^*\}\xrightarrow{\text{PIC($N^*$ clusters)}}\{z_1^*,...,z_{N_r}^*\}$
 \caption{SSC algorithm for joint representation learning and
clustering}
\label{fig:ssc_algo}
\end{algorithm}

\subsection{DNN Training - Dynamic Triplet Similarity}\label{sec:triplet_loss}
The clustering algorithm generates labels $\bm{z}^{q-1}$ where each $z^{q-1}_i$ can take discrete values from  $\{1,..,N_{q-1}\}$, with $N_{q-1}$ being the number of clusters at the end of iteration ($q-1$). We use the dynamic triplet similarity to train the DNN model. The term dynamic is used to indicate  the formation of new triplets after each iteration. We form a positive pair $\{\textbf{y}^{q-1}_i,\textbf{y}^{q-1}_j\}$ by selecting the anchor $\textbf{y}^{q-1}_i$ and positive sample $\textbf{y}^{q-1}_j$  from each cluster $\mathcal{C}^{q-1}_a$. Here, $\textbf{y}^{q-1}$ denotes the representations formed at the DNN output using $\boldsymbol{\theta}^{q-1}$.  The negative pair is formed by randomly sampling the embedding $\textbf{y}^{q-1}_l$ from any other cluster $\mathcal{C}^{q-1}_b$ ($a \neq b$). The DNN model parameters $\boldsymbol{\theta}$ are updated based on the following optimization criteria. 
\begin{equation}
\vspace{-8pt}
\boldsymbol{\theta}^q =   \underset{\boldsymbol{\theta}}{\operatorname{argmax}}\sum_{i, j, l} \big [ s(i,j) - { \alpha \left( s(i,l)+s(j,l) \right)}\big ]
\label{eq:triplet_loss}
\end{equation}
where, $s(i,j)$ is pairwise similarity score between $\bm{y}^{q-1}_i$ and $\bm{y}^{q-1}_j$. Here, $0<\alpha\leq1$ is a weighting factor which controls the discriminability of the negative pairs. In the triplet formation, we also try to uniformly sample similar number of anchors from all the clusters by suitably oversampling the negative pairs for the same positive pair. 

\subsection{SSC Algorithm Overview}
The overview of the SSC\footnote{Our implementation of SSC-PIC is available at \url{https://github.com/iiscleap/SSC.git}} algorithm is given in Algorithm \ref{fig:ssc_algo}. The DNN model is trained using gradient ascent with triplet similarity. At the $q$-th iteration, positive and negative triplet pairs are formed using the cluster label indices $\textbf{z}^{q-1}$ of the previous iteration. We calculate the triplet similarity using these triplets and learn the DNN model parameters $\boldsymbol{\theta}^q$ with the inputs as x-vector features $\bm{X}_r$.  


\begin{algorithm}[t!]
\SetAlgoLined
\begin{enumerate}
\setlength\itemsep{0.5em}
  \item $\Y^q\xrightarrow{\text{PIC($N^{q-1}$ clusters})} \boldsymbol{A} \in \mathbb{R}^{N^{q-1}XN^{q-1}}:$ Affinity matrix of $N^{q-1}$ clusters such that $[A]_{ij}=\mathcal{A}(\mathcal{C}_i,\mathcal{C}_j)$
\item $\lambda_1\geq\lambda_2\geq...\geq\lambda_{N^{q-1}}:$ Eigen values of $\A$;\\ $total\_energy=\sum_{i=1}^{N^{q-1}}\lambda_i$
\item $\boldsymbol{v} \in \mathbb{R}^{N^{q-1}}:$ cumulative explained variance ratio;\\
$[v]_k=\frac{\sum_{i=1}^{k}\lambda_j}{total\_energy}$ where $k \in \{1,...,N^{q-1}\}$
\item $N^q=\argmax\limits_k ([v]_k\leq \phi^q)$
\item return $N^q$
\end{enumerate}
\caption{Estimate number of clusters $N^q$: Estimate$\_N^q(\Y^q,\phi^q,N^{q-1})$ }
\label{fig:estimate}
\end{algorithm}

Once the DNN model is trained, we obtain embeddings $\Y^q$ from the last layer of DNN. We compute cosine similarity scores matrix $\bm{S}$ using the embeddings $\Y^q$ for the recording. 

The PIC based clustering is performed using the similarity measure such that $N^{q} \leq N^{q-1}$. We estimate $N^q$ based on the stopping threshold $\phi^q$ discussed in next section. We start with $N_r$ clusters and perform multiple merges using the criterion defined in Equation~(\ref{eq:ahc_cluster_affinity}) thereby reducing the total number of clusters in the $q$-th iteration. If $N^{q}=N^*$ or the stopping criterion is met, then the algorithm is stopped. We perform one more iteration of DNN training and clustering to generate the final diarization results. Else, the iteration index is updated and the steps described above are repeated for the $(q+1)$-th iteration. If $N^*$ is unknown, then the stopping criteria is based on total number of iterations $Q$ or till we reach $N^q=1$ (whichever is achieved first). 

\subsection{Estimation of Number of Clusters (Algorithm \ref{fig:estimate})}\label{sec:stopping_criterion}
The number of clusters $N^q$ required for clustering at each iteration $q$ is estimated based on explained variance. In this approach, we compute affinity matrix $\boldsymbol{A}$ of $N^{q-1}$ clusters using Equation~(\ref{eq:affnty_pic}). The diagonal elements are set as the maximum value of non-diagonal elements of the matrix. Then, we compute eigen values of the matrix and arrange them in descending order. We accumulate the eigen values till the cumulative explained variance ratio, which is the ratio of accumulated eigen values to total sum of eigen values, reaches a stopping threshold $\phi^q$ for the $q$-th step. The number of accumulated eigen-values at this step is denoted as $N^q$.

\subsection{Incorporating Temporal Continuity}
In the given audio stream, the event that two neighboring x-vector embeddings come from the same speaker is more likely than the event that they belong to different speakers. This heuristic can be incorporated in the clustering algorithm. After DNN training, we obtain similarity score matrix $\bm{S}$ using the output embeddings and multiply the similarity score  $s({i,j})$ with an exponential decaying function of the temporal distance between $i$ and $j$. 
\begin{equation} 
    s'(i,j) = s(i,j)\beta^{min(n_b,|i-j|)}
    \label{eq:temp_continuity}
\end{equation}
where, $\beta$ is a positive decaying factor ($0<\beta<1$), $|i-j|$ is the absolute value of the time difference between segment $i$ and $j$, $n_b$ denotes a floor value which prevents the similarity score from going too low for segments that are farther in time.  The above modification of similarity score encourages neighboring clusters to merge and have smooth transition among clusters.

\section{Experimental Setup}\label{sec:data}
\subsection{Evaluation Data}
\begin{itemize}
    \item CALLHOME: The CALLHOME (CH) dataset~\cite{callhome} is a collection of multi-lingual telephone data sampled at $8$kHz, containing $500$ recordings, where the duration of each recording ranges from $2$-$5$ mins. The number of speakers in each recording varies from $2$ to $7$, with majority of the files having $2$ speakers.   The CH  dataset is  divided equally into $2$ different sets, CH1 and CH2, with similar distribution of number of speakers. 
    \item AMI: The AMI dataset~\cite{mccowan2005ami} comprises of meetings recorded at different sites (Edinburgh, Idiap, TNO, Brno) with sampling frequency of $16$kHz. We use the dev and eval sets of the single distant microphone (SDM) condition from the ofﬁcial speech recognition partition of AMI dataset.
    The AMI-dev set contains $18$ meeting recordings and AMI-eval set contains $16$ meeting recordings. All the meeting recordings contain $4$ speakers except one recording from the  eval set which has $3$ speakers. The duration of each recording ranges from $20$-$60$ mins. 
\end{itemize}

\begin{table}[t!]
\centering
\caption{The x-vector training configurations for baseline of CH and AMI datasets. }
\label{tab:comparision_xvec}
\begin{tabular}{lll}
\hline
{
\textbf{Parameter}}           & {\textbf{CH}} & {\textbf{AMI} }   \\ \hline
{Sampling Rate  }              & {8kHz }       & {16kHz }          \\
{Train set }                   & {SWBD, SRE 04-08 }   & {Voxceleb 1,2} \\
{\# speakers } & {4,285}       & {7,323 }          \\
{Input features}               & {23D MFCCs}  & {30D MFCCs}      \\
{X-vector dimension}           & {128}         &{ 512 }            \\ \hline

\end{tabular}
\end{table}

\subsection{X-vector Extraction} 
{The details of the x-vector model are given in Table~\ref{tab:comparision_xvec}}. 
The TDNN model consists of $5$ layers of time-delay neural network and two layers of feed forward architecture operating at frame-level followed by a pooling layer which generates mean and standard deviation statistics at the segment level \cite{snyder2018x}. The segment level statistics are passed through two feed forward layers to the target layer. The model is trained using cross-entropy loss on the training speakers using the asynchronous stochastic gradient descent (ASGD) algorithm.   The first affine layer (before the non-linearity) after the pooling layer is used as the x-vector feature. Table \ref{tab:comparision_xvec} shows the differences in x-vector training configuration for CH and AMI datasets.

\subsection{Experiments on CH Dataset}\label{sec:CH}
\subsubsection{Baseline}\label{sec:baseline_ch} 
The baseline system used for the  CH diarization task comes from the Kaldi recipe\footnote{\url{https://kaldi-asr.org/models/m6}}. It involves the extraction of $23$ dimensional mel-frequency cepstral coefﬁcients (MFCCs). A sliding window mean normalization is applied over a $3$s window after the feature extraction. 
For training the x-vector TDNN model and the PLDA model, we use the SRE 04-08 and Switchboard cellular datasets as given in the Kaldi recipe~\cite{povey2011kaldi}. This training set had $4,285$ speakers. For the diarization task,  speech segments ($1.5$s chunks with $0.75$s overlap) are converted to $128$ dimensional  neural embeddings (x-vectors)~\cite{snyder2018x}. The x-vectors are processed by applying whitening transform obtained from the held-out set followed by length normalization~\cite{garcia2011analysis}. An utterance level PCA is applied for dimensionality reduction~\cite{zhu2016online} preserving  $10$\% of total energy. These embeddings are used in the PLDA model to compute the similarity score matrix. 
The clustering is performed using the AHC algorithm as discussed in Section \ref{sec:AHC}. The AHC stopping criterion is determined using held-out data (CH1 is used for diarization on CH2 data and vice-versa). 

\subsubsection{SSC Algorithm Implementation}
We use the same setup in the baseline system to extract the x-vector features. The number of x-vectors per recording ($N_r$) range from $50$-$700$.
The DNN model is a two-layer fully connected DNN. The first layer has $128$ input and hidden nodes. The second layer has $10$ output nodes.  
The first layer of the DNN also realizes the unit length normalization as the non-linearity.   The learning rate is set at $0.001$. The model is trained using the triplet similarity defined in Equation~(\ref{eq:triplet_loss}). We do not split the training triplet samples into mini-batches, but rather perform a full batch training with Adam optimizer~\cite{kingma2014adam}. We use a parameter $\eta=0.5$ (defined as the fraction of the training similarity measure at the zeroth iteration compared to the current iteration) for early-stopping of the DNN model training. Typically, this model training involves $5$-$10$ epochs. The DNN outputs ($10$ dimensional embeddings) are used to compute the pairwise similarity score matrix using cosine similarity for the PIC.

\subsection{Experiments on AMI Dataset}\label{sec:AMI}
\subsubsection{Baseline}\label{sec:baseline} 
The baseline x-vector system uses the DIHARD Challenge recipe\footnote{\url{https://github.com/iiscleap/DIHARD_2019_baseline_alltracks}} for the AMI dataset. It involves the  extraction of $30$ dimensional MFCCs. The x-vectors are of dimension $512$. For training the x-vector extractor and the PLDA model in AMI experiments, we trained the TDNN model using $16$ kHz data from  VoxCeleb-1 \cite{nagrani2017voxceleb} and VoxCeleb-2 \cite{nagrani2020voxceleb} datasets containing $7,323$ speakers. In the post-processing of the x-vectors, we use the held-out dataset from the second DIHARD challenge~\cite{ryant2019second} for computing the whitening transform. The rest of the steps in the PLDA-AHC baseline system follow the same processing pipeline used in the CH baseline system.  
        
\begin{figure}[t!]
	\centering
	\includegraphics[trim={0cm 0cm 0cm 0cm},clip,width=\linewidth]{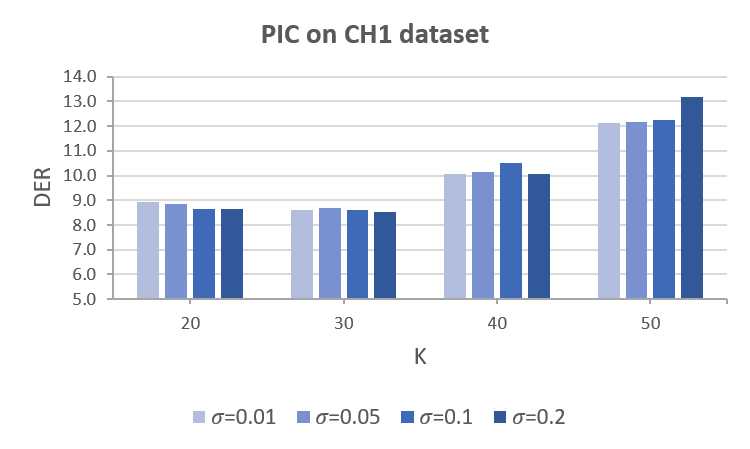}
	\caption{DER of CH1 dataset for different choices of number of  nearest-neighbors $K$ and scaling factor $\sigma$.}
	\label{fig:K_z_plot}
	\vspace{-10pt}
\end{figure}

\begin{figure}[t!]
    \centering
    \includegraphics[trim={0cm 0cm 0cm 0cm},clip,width=0.8\linewidth]{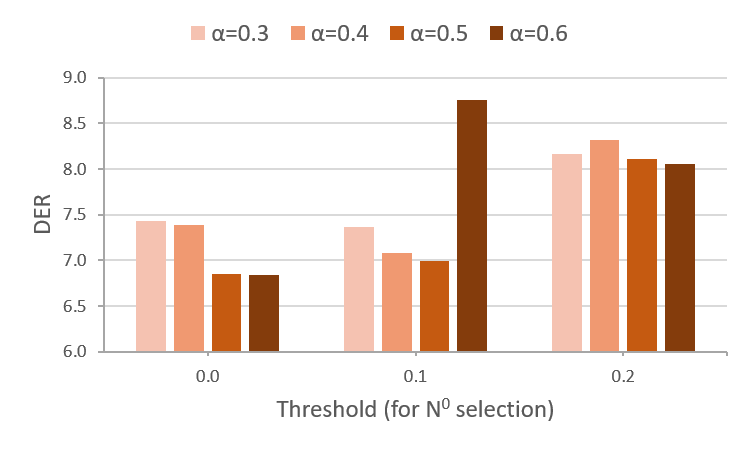}
    \caption{Bar plot of DER vs threshold for $N^0$ selection on CH1 subset of CALLHOME for different choices of $\alpha$ parameter.
    }
    \label{fig:th_alpha}
\end{figure}

\begin{figure}[t!]
    \centering
    \includegraphics[trim={0cm 0cm 0cm 0cm},clip,width=0.8\linewidth]{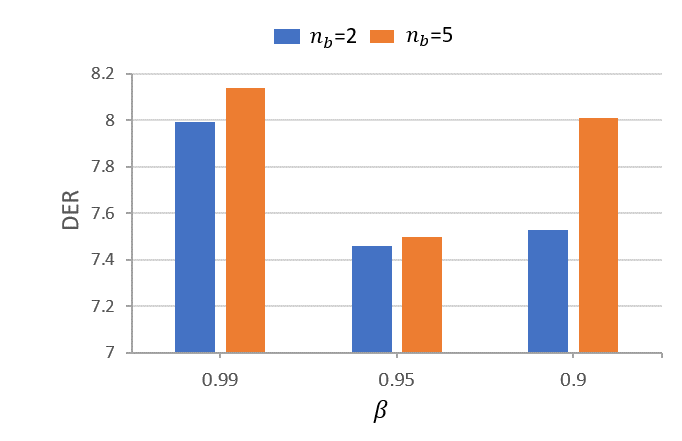}
    \caption{Effect of temporal weighting ($\beta, n_b$) on DER for CH1 dataset. The best DER is obtained for  $n_b=2$ with $\beta = 0.95$.}
    \label{fig:temporal}
\end{figure}
\subsubsection{SSC Algorithm Implementation}\label{sec:ami_spec}
The DNN model is a two-layer feed forward architecture (similar to the CH dataset). The first layer has $512$ input and hidden nodes. The second layer has $30$ output nodes (similar to the PCA dimension used in the baseline system).  In the AMI dataset, the number of x-vectors per recording range from $1000$-$4000$. Hence, a large number of triplets can be generated  for each recording. We  resort to the use of minibatches in DNN model training. A validation split is also formed in each recording. The loss on the  validation set is used to decide the stopping point for training. We perform a learning rate annealing~\cite{Lin2019} in the DNN model training.  The initialization of the DNN model follows the strategy used in CH dataset. 
\begin{figure*}
\centering
 \includegraphics[trim={0cm 4cm 0cm 0cm},clip,width=0.9\textwidth]{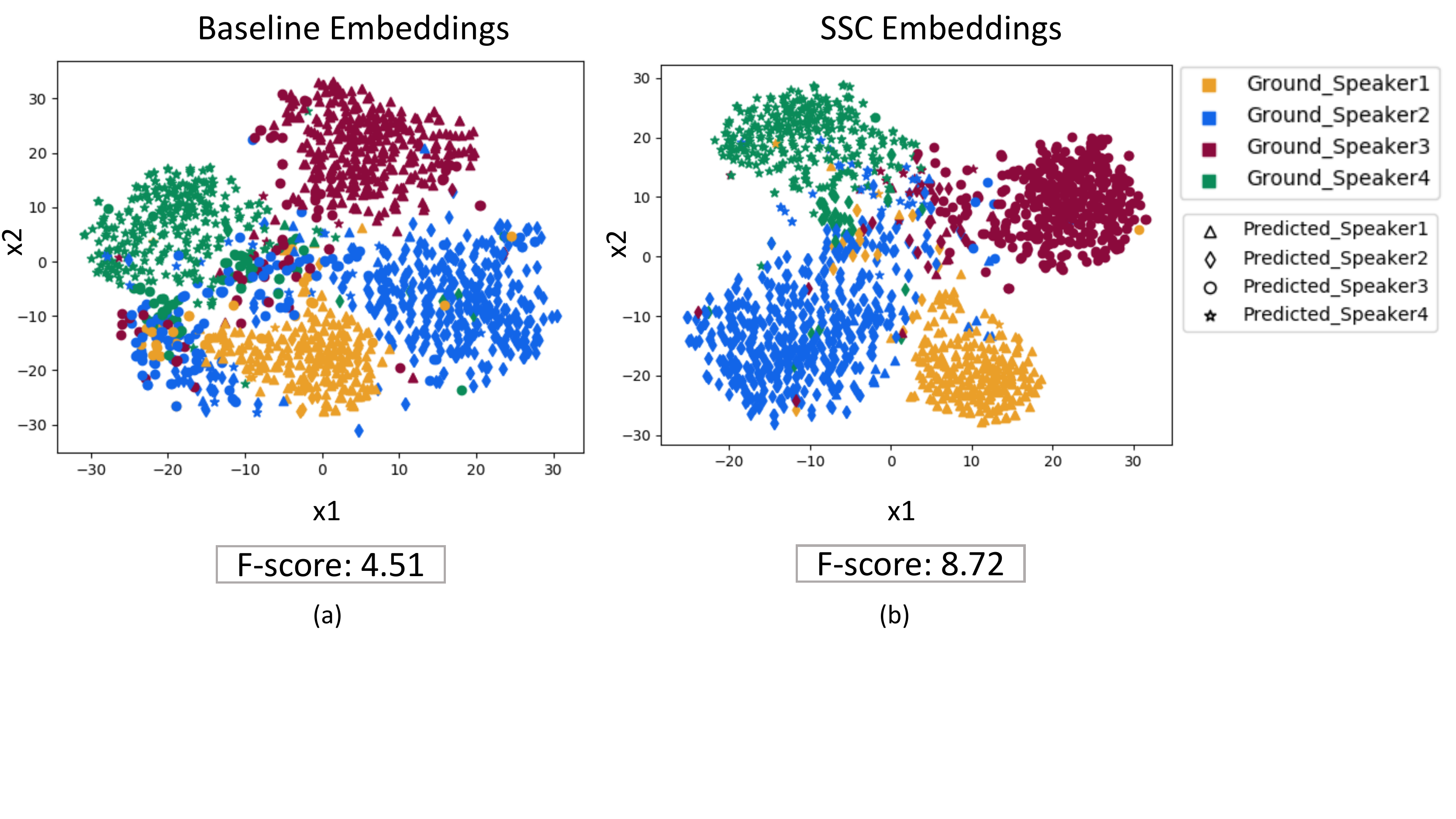}
 \vspace{-0.25cm} 
  \caption{t-SNE based visualization of embeddings extracted on $1.5$s audio segments  from the recording {AMI-ES2011c-SDM}. (a) the baseline x-vectors post-processed with whitening transformation and PCA, (b) embeddings obtained after SSC algorithm (DNN embeddings $\Y $).  The colors represent speakers in the ground truth and the shapes represent predicted clusters. }
    \label{fig:tsne}
\end{figure*}

\section{Results and Analysis}\label{sec:analysis}
The performance metric in all the experiments used in this paper is the diarization error rate (DER) computed with a $250$ ms collar and ignores overlapping segments. For all our experiments, we use the oracle speech activity decisions. 
\subsection{Choice of Hyper-parameters}

\begin{itemize}[leftmargin=*,listparindent=\parindent]
     \item \textbf{Number of nearest-neighbours $K$ and scaling factor $\sigma$:} We choose $K$ and $\sigma$ values based on experiments on the CH1 subset of the CALLHOME dataset. Figure \ref{fig:K_z_plot} shows the  overall DER on CH1 for different values of $K$ and $\sigma$. We observe that as we increase $K$, the DER is increasing. The higher $K$ value also increases computational complexity. Based on the results, the value of $K$ is chosen as $30$. For a fixed $K$, we vary $\sigma$ from $0.01$ to $0.2$. The parameter $\sigma$ has only a minor impact on the final DER performance. The value $\sigma=0.1$ is chosen for all our experiments. 

    \item \textbf{Initial number of clusters ($N^0$) and $\alpha$:}  The DNN training is performed only after the PIC/AHC clustering algorithm is performed for a few iterations to reduce the number of clusters from $N_r$ to $N^0$. 
    If the value of $N^0$ is too large, then the discriminative triplet similarity measure tends to increase the  within speaker diversity (controlled by factor $\alpha$). However, if the value of $N^0$ is reduced significantly, then the clusters formed can be potentially impure in terms of the speaker identity of the elements in the cluster. 
    
    For CH dataset, we initialize the model using AHC and vary the threshold applied on similarity scores to stop at $N^0$ number of clusters. In Figure \ref{fig:th_alpha}, we observe that for lower threshold (smaller $N^0$ ), higher $\alpha$ (more discriminative) is better but as we increase threshold (larger $N^0$), lower $\alpha$ (less discriminative) is optimal. We obtain the best DER with threshold$=0.0$ and $\alpha=0.6$, which is used in CH experiments. In the AMI experiments, we choose $N^0$ as the smallest number obtained using the threshold $\phi$ for the explained variance on the affinity.
    
    \item \textbf{Temporal weighting factors $\beta$ and $n_b$:} Figure~\ref{fig:temporal} shows the effect of the hyper-parameters $\beta$ and $n_b$ used in incorporating temporal continuity   (Equation~\ref{eq:temp_continuity}). The best DER is obtained on the CH1 dataset using $n_b =2 $ and $\beta=0.95$. The inclusion of temporal continuity discourages frequent speaker turns. Therefore, it is more useful for recordings which are longer in duration and for those with less frequent speaker turns. We found the best parameters on  CH1 subset and used it for both the CH and AMI experiments.
\end{itemize}
{
\subsection{ Triplet sampling strategy}
 We explore different sampling strategies and measure DER performance on CH1 dataset:
\begin{enumerate}
    \item \textbf{Hard Sampling}: We sample positive pairs from the same cluster and negative from the nearest neighbor of the anchor belonging to a different cluster. DER obtained on CH1 is 9.4\%.  
    \item \textbf{Random Sampling}: Here, we sample positive pairs from the same cluster and randomly sample negative pairs from any other cluster. We get DER$=6.8\%$.
    \item \textbf{Easy Sampling}: In this case, we sample positive pairs from the same cluster and the negative pair using the farthest data point from the anchor that belongs to a different cluster. 
    DER performance is $8.0\%$.
\end{enumerate}
Although hard sampling is typically preferred in triplet mining on supervised data, we find that the random sampling works best in our case of unsupervised clusters.  The hard sampling may potentially bias the model to disallow the merge of same speaker clusters. However in case of easy sampling, the SSC model learning is somewhat compromised as the negative samples do not provide a discriminative learning criterion. 
}
\subsection{SSC Initialization}
{  The initialization of the DNN model uses the whitening transform and the recording level PCA from the baseline system for the first and the second layer respectively. The output of the model  are the embeddings $\Y=\{\bm{y}_1,..,\bm{y}_{N_r}\}$ which are used to perform clustering (AHC/PIC) till $N^0$ number of clusters.   For CH, we use AHC clustering with threshold$=0.0$ to generate $N^0$ cluster labels for each $\bm{y}_k$. For the AMI SSC-AHC training, we used PLDA scoring with learnable PLDA parameters. 
We use PLDA-AHC with threshold$=0.0$ to generate initial speaker labels .
For SSC-PIC, we perform PIC clustering using either $N^0=N^*$ in the known $N^*$ case  or we generate $N^0$ using a stopping threshold $\phi=0.7$ on the eigen value ratio (Algorithm \ref{fig:estimate}) in the unknown $N^*$ case.}

{
\subsection{Stopping Criterion}
In the SSC training, the stopping criterion for the DNN representation learning model is based on the triplet-loss. We use the stopping threshold of $0.5$ times the triplet-loss at the initial epoch of the model training.     The stopping criterion for the SSC merging process is  done using the eigen-value ratio  in the affinity matrix (Sec.~\ref{sec:stopping_criterion}). The exact threshold is selected based on the development data (similar to the AHC threshold estimation).  }

\subsection{Visualization of Embeddings}
Figure~\ref{fig:tsne} shows the visualization of the embeddings from the baseline system (x-vec. + AHC) and the proposed SSC model for one recording from the AMI dataset. The visualization is performed using t-distributed  stochastic  neighborhood embedding (t-SNE) which performs an unsupervised dimensionality reduction of the embeddings \cite{maaten2008visualizing}. The embeddings from the baseline system are shown in Figure~\ref{fig:tsne}(a) where the x-vectors are extracted from $1.5$s segments of audio with half overlap followed by a whitening transform and PCA dimensionality reduction at the recording level. The embeddings from the SSC algorithm are shown in   Figure~\ref{fig:tsne}(b)  for the same recordings. The colors and shapes indicate the ground-truth and predicted speaker clusters respectively. The ideal embeddings would have  one-to-one mapping between shapes and colors. As seen in this Figure, the SSC algorithm provides improved cluster separability (highlighted using F-score in the plots) along with improved  association with the ground-truth speakers.

\begin{table}[t!]
\centering 
\caption{\color{black} DER (\%) for different systems when number of speakers ($N^*$) is known and unknown for the full CH dataset. }
\label{tab:callhome}
\begin{tabular}{|l|c|c|}
\hline
\textbf{System}                                                                             & \textbf{Known N*} & \textbf{Unknown N*} \\ \hline
x-vec. + cosine + AHC                                                                       & 8.9        & 10.0           \\
x-vec. + cosine + Spec. Clus.                                                                        & 9.4         & 11.9          \\
x-vec. + PLDA + AHC (Baseline)                                                              & 7.0         & 8.0           \\ \hline \hline 
x-vec. + cosine + PIC                                                                       & \color{black}7.7         & \color{black}9.3                \\  
Self-supervised AHC (SSC-AHC) \cite{Singh2020}    & \multirow{1}{*}{6.4}         & \multirow{1}{*}{8.3}           \\
Self-supervised PIC (SSC-PIC) & \multirow{1}{*}{\color{black}6.4} & \multirow{1}{*}{7.5} \\ 
+ Temporal continuity &  \textbf{\color{black}6.3} &  \textbf{7.0}              \\ 
                            \hline
\end{tabular}
\vspace{-0.2cm}
\end{table}
\subsection{CH Diarization Results}
The diarization results on the CH dataset are shown in Table~\ref{tab:callhome}. Here, the initial set of experiments use x-vector embeddings (with post-processing) clustered with different algorithms (AHC/PIC/Spectral Clustering) and similarity measures (cosine/PLDA). { The spectral clustering is a graph partitioning algorithm which uses eigen values and eigen vectors of Laplacian matrix to perform clustering~\cite{park2019auto}. The Laplacian is computed from similarity score matrix. We use the eigen vectors with non-zero eigen values to perform k-means clustering.} 

{The PLDA-AHC approach gives the best performance for x-vector embeddings and this is used as the baseline system for comparison with the proposed SSC algorithm. For known $N^*$, Self-Supervised PIC (SSC-PIC) shows marginal improvement compared to SSC-AHC. However, for unknown $N^*$, we obtain considerable improvements for the SSC-PIC algorithm. The best results improve significantly over  the baseline system for both the cases of known and unknown number of speakers.} The relative improvements in known $N^*$ and unknown $N^*$ respectively for the SSC algorithm over the baseline system are $10$\% and  $13$\%. { We also performed paired student-t test to check the statistical significance of file-level DER improvements obtained for the SSC-PIC over the baseline system  in Table \ref{tab:callhome}. In this statistical t-test, comparing the baseline and SSC-PIC system, we find a t-statistic  of $t=2.2$ and p value of  $p=0.03$. This indicates that the improvements seen in the proposed approach are statistically significant. In order to show the improvement in per-cluster variance, we computed the F-ratio (ratio of between speaker variance to within speaker variance) based on the similarity scores. We found the F-ratio to improve by $130$\% on an average  on the CH1 dataset with the SSC training.}



\begin{table}[t!]
\caption{\color{black}{DER (\%) for different systems when number of speakers ($N^*$) is known and unknown for the AMI dataset.}}
\label{tab:ami_known}
\centering
\begin{tabular}{|l|c|c|c|c|} 
\hline

\multirow{2}{*}{\textbf{System}} & \multicolumn{2}{c|}{\textbf{Known $N^*$}}& \multicolumn{2}{c|}{\textbf{Unknown $N^*$}} \\ 
\cline{2-5}
& Dev.   & Eval. & Dev. & Eval.  \\ 
\hline\hline

x-vec + cosine + AHC  & 34.6  & 30.2        &18.2 & 15.5            \\ 
x-vec + cosine + Spec. Clus.                                                                                              & 30.2 & 25.5     & 40.0 & 31.1               \\ 
x-vec + PLDA + AHC (Baseline)                                                                                    & 15.7 & 16.0        & 13.7 & 16.3            \\ 

\hline \hline
 SSC-PLDA-AHC               & 9.4 & 11.1     &  10.7   &   11.6             \\
x-vec + PLDA + PIC   & \color{black}9.4   & \color{black} 9.3 & \color{black}9.8 & \color{black}10.4                    \\
x-vec + cosine + PIC   & \color{black}8.9 & \color{black} 7.3 & \color{black}9.0 & \color{black}7.3                     \\ 
SSC-PIC               & \color{black}7.3 & \color{black}7.2     & \color{black} 8.1   & \color{black}7.6               \\
+ Temporal continuity & \textbf{\color{black}6.2}  & \textbf{\color{black}6.4}    & \textbf{\color{black}6.4} & \textbf{\color{black}6.7}            \\
\hline
\end{tabular}
\end{table}

\subsection{AMI Diarization Results}
The diarization results on the AMI dataset are shown in Table~\ref{tab:ami_known}. Here, the initial set of experiments report using x-vector embeddings (with post-processing) clustered with different algorithms (AHC/PIC/Spec. Clus.) and similarity measures (cosine/PLDA). {As the PLDA-AHC performance is significantly better than cosine-AHC, we performed SSC-AHC with PLDA scoring.  Our SSC-PLDA-AHC system outperforms the  baseline for all scenarios. However, the PIC algorithm achieves the best results compared to other clustering results on AMI. PIC algorithm can be used with PLDA and cosine scoring but the performance is comparatively better with cosine because the cosine scores are bounded which leads to uniformity while clustering.} The cosine scoring with PIC gives $43$\% and $54$\% relative improvements for dev and eval respectively for known $N^*$ over the baseline system. The SSC algorithm performed with PIC provides additional gains. 
 The incorporation of temporal continuity also improves results. We obtain significant relative improvements ( $60$\% for both dev and eval ) for known $N^*$. For unknown $N^*$, we obtain relative improvement of $53$\% and $59$\% for dev and eval respectively over the baseline system.

\subsection{Comparison With Other Published Works}
Prior works on AMI report results on three domains {(Edinburgh, Idiap and Brno)} out of the four domains. For {comparison}, we have also shown SSC-PIC results for these three domains in Table~\ref{tab:comparision_AMI}. As seen here, the proposed SSC algorithm advances the {state-of-the-art} results significantly for both the known/unknown $N^*$ condition. 

A similar comparison with other published results on the CH dataset is shown in Table~\ref{tab:comparision}. A much larger pool of published results exist on the CH dataset in the recent years. Based on our survey of recent literature, the best reported results use the eigen-gap based spectral clustering \cite{park2019auto}.  The SSC approach  improves over {state-of-the-art} in CH dataset results as well. { However, improvements on AMI dataset shows more significant improvements compared to CH dataset with PIC. One major reason is the difference in the duration of recordings between the datasets. In PIC, the graph is more stable when there are more nodes to build the edge connections. The duration of AMI recordings ranges from $20$-$60$ mins which generates $1000$-$4000$ embeddings. In the CH dataset, the recording duration ranges from $2$ -$5$mins generating $50$-$400$ embeddings. Further, the SSC training also benefits from longer duration as we have more triplets ($>200,000$) from AMI to train the network as opposed to CH dataset with $<50,000$ triplets.}

The SSC algorithm performance is also shown to be superior to the fully supervised diarization algorithms (for example, the RNN based algorithm~\cite{zhang2019fully}) which use a significantly large  amount of speaker supervised conversational data. In contrast, the proposed SSC algorithm is purely unsupervised and it is developed without any additional training data over the baseline system. Table~\ref{tab:comparision_AMI} and \ref{tab:comparision} therefore highlight the advantages  of self-supervised learning and graph structural clustering used in the SSC algorithm. The SSC outputs at segment level can also be used as initialization for further refinement using frame-level VB-HMM modeling~\cite{diez2018speaker}. As seen in Table~\ref{tab:comparision}, this VB-HMM based refinement step further improves the diarization performance to achieve a DER of $4.8$\% for known $N^*$ case and $5.6$\% for unknown $N^*$ case on the CH dataset.



\begin{table}[t!]
\centering
\caption{Comparison of the proposed SSC algorithm  with other published works on the AMI dataset (after removing recordings from TNO domain as reported in other works). }
\label{tab:comparision_AMI}
\begin{tabular}{|l|c|c|c|c|} 
\hline
\textbf{System}                      & \multicolumn{2}{c|}{\textbf{DER known N*}} & \multicolumn{2}{c|}{\textbf{DER unknown N*}}  \\ 
\cline{2-5}
                                     & Dev.   & Eval.                               & Dev.   & Eval.                                  \\ 
\hhline{|=====|}
Semi-sup learning \cite{pal2019study} & - & - & 17.5 & 22.0 \\
Incremental learning \cite{dawalatabad2019incremental} & - & 15.6 & - & 20.0 \\
 GAN clustering~\cite{pal2020speaker} & 10.2 & 10.1                               & 11.0 & 11.3                                 \\ 

2-D self attention \cite{sun2019speaker} & ~~~ - & ~~ -                               & 12.2 & 13.0                                 \\ 
\hline 
Baseline    & 14.4 & 16.5                               & 12.9 & 13.6                                 \\ 
SSC-PIC                                  & \textbf{{\color{black}4.6}}  & \textbf{{\color{black}6.5}}                                & \textbf{{\color{black}5.2}}  & \textbf{{\color{black}5.4}}                                  \\
\hline
\end{tabular}
\end{table}
\begin{table}[t!]
	\centering
	\caption{Comparison of the proposed SSC algorithm based results with other published works on the CH  dataset. Here, VB refers to the use of VB-HMM based refinement.}
	\label{tab:comparision}
	\begin{tabular}{|l|c|c|} 
		\hline
		\textbf{System}  & \textbf{DER Known N*} &\textbf{DER unknown N*}  \\ \hline 
		x-vec \cite{garcia2017speaker} (+VB)  & - & 12.8 (9.9) \\
		VB modeling \cite{diez2018speaker}  & - & 9.0 \\ 
		Bootstrap network \cite{li2019discriminative}  & 6.2 & 8.6 \\ 
		LSTM scoring \cite{Lin2019} & - & 7.7 \\
		UIS-RNN~\cite{zhang2019fully}  & - & 7.6 \\
		Spec. Cluster \cite{park2019auto}  & - & 7.3 \\ 
		\hline 
		Baseline (+VB)  & 7.0 (5.0) & 8.0 (6.4) \\
		SSC-PIC (+VB)  & \textbf{\color{black}6.3} (\textbf{4.8}) & \textbf{7.0} (\textbf{5.6}) \\
		\hline 
	\end{tabular}
	\vspace {-0.3cm} 
\end{table}
\subsection{Computational Complexity}
The training of DNN and the PIC are two stages to be considered for time complexity at recording level. Based on the triplet sampling strategy  discussed in Section \ref{sec:triplet_loss}, the time complexity for training the DNN at each epoch is $\mathcal{O}(N_r^2)$. The time complexity for clustering stage is based on choice of algorithm. The baseline AHC has $\mathcal{O}(N_r^3)$ complexity. The proposed PIC algorithm is more efficient in computation as it calculates incremental path integral in linear time and does not require re-computation of similarity scores. It has time complexity of $\mathcal{O}(N_r^2)$. The overall time complexity of SSC-PIC is $\mathcal{O}(RQN^2)$, where $R$ is the number of DNN epochs and $Q$ is the number of SSC iterations performed. Comparing with the baseline system complexity of    $\mathcal{O}(N_r^3)$, the proposed SSC algorithm is more efficient when $N_r > RQ$. This condition is satisfied for recordings that are  $30$s or more in duration.  We performed a compute-time measurement for the baseline (PLDA + AHC) and SSC-PIC system using x-vectors on an Intel CPU  server (single core 64-bit).  The compute time for the baseline and proposed SSC were $71.6s$ and $43.3s$ respectively for a file of duration $26$ minutes from the AMI dev set. 
This shows that the SSC algorithm achieves an improved DER performance while also reducing the computational complexity.

\section{Conclusion}\label{sec:concl}
In this paper, we have proposed an algorithm for self-supervised embedding learning and clustering based on graph-structural path integral. The steps of embedding learning and path integral clustering (PIC) are performed iteratively for the given recording using x-vector features as inputs. The iterative steps validate the hypothesis that improved clustering can be achieved using discriminative representations while self-supervised representations can be learnt with well-separated clusters.  The proposed approach, termed as self-supervised clustering (SSC), is applied for speaker diarization task in CH and AMI datasets. The diarization error rates achieved in these tasks are shown to improve the baseline system as well as several other recent {state-of-the-art} techniques proposed in the field.  To the best of our knowledge, the DER results reported in this work constitute the lowest error rates reported till date for diarization on CH and AMI datasets.  
\section*{Appendix A - Proof of Path Integral}\label{appendix:proof}

The unnormalized path in Equation \ref{eq:path_integral} is:
\begin{eqnarray}
g_{i j}&=&\delta_{i j}+\sum_{k=1}^{\infty} \sigma^{k} \sum_{\gamma \in \Gamma_{i j}^{(k)}} \prod_{s=1}^{k} p_{u_{s-1}, u_{s}}\nonumber\\
&=&\delta_{i j}+\sum_{k=1}^{\infty}\sigma^{k}[\bm{P}_{\mathcal{C}_{a}}^k]_{ij}\nonumber\\
&=&[\boldsymbol{I} +\sum_{k=1}^{\infty}\sigma^{k}\bm{P}_{\mathcal{C}_{a}}^k]_{ij}\nonumber \\
&=&[(\boldsymbol{I} - \sigma\bm{P}_{\mathcal{C}_{a}})^{-1}]_{ij}\nonumber
\end{eqnarray}
The normalized path integral $\mathcal{S}_{\mathcal{C}_{a}}$ (Equation \ref{eq:pic_ca}) is given as,
\begin{eqnarray}
[\mathcal{S}_{\mathcal{C}_{a}}]_{i j}&=&\frac{1}{|\mathcal{C}_a|^2}[(\boldsymbol{I} - \sigma\bm{P}_{\mathcal{C}_{a}})^{-1}]_{ij}\nonumber
\end{eqnarray}




\ifCLASSOPTIONcaptionsoff
  \newpage
\fi



\bibliographystyle{IEEEtran}
\bibliography{main}
%



%

\begin{IEEEbiography}[{\includegraphics[trim={0cm 0cm 0cm 0cm},width=1in,height=1.25in,clip,keepaspectratio]{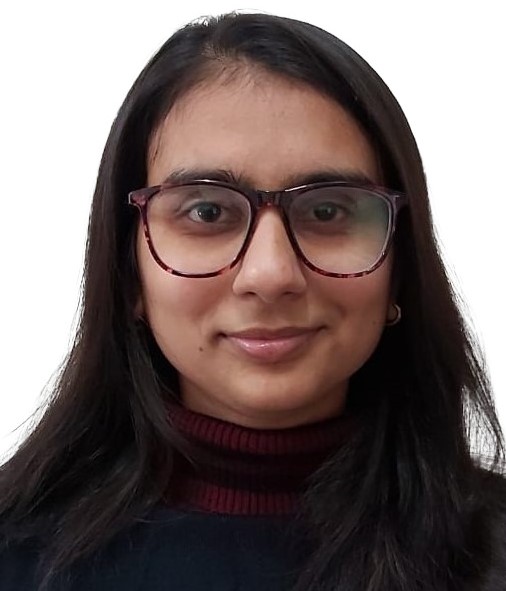}}]{Prachi Singh}
is a graduate student at Learning and Extraction of Acoustic Patterns (LEAP) lab, Electrical Engineering, Indian Institute of Science, Bangalore. In the past, she has  worked in Fiat Chrysler Automobiles, Chennai, India from 2015 to 2017. She obtained her Bachelor of Technology in Electronics and Telecommunication from College  of  Engineering,  Pune in  2015. Her research interests include speaker diarization, speaker verification, self-supervised learning and graph clustering. She is a student member of IEEE and ISCA. 
 \end{IEEEbiography}
 \begin{IEEEbiography}[{\includegraphics[width=1in,height=1.25in,clip,keepaspectratio]{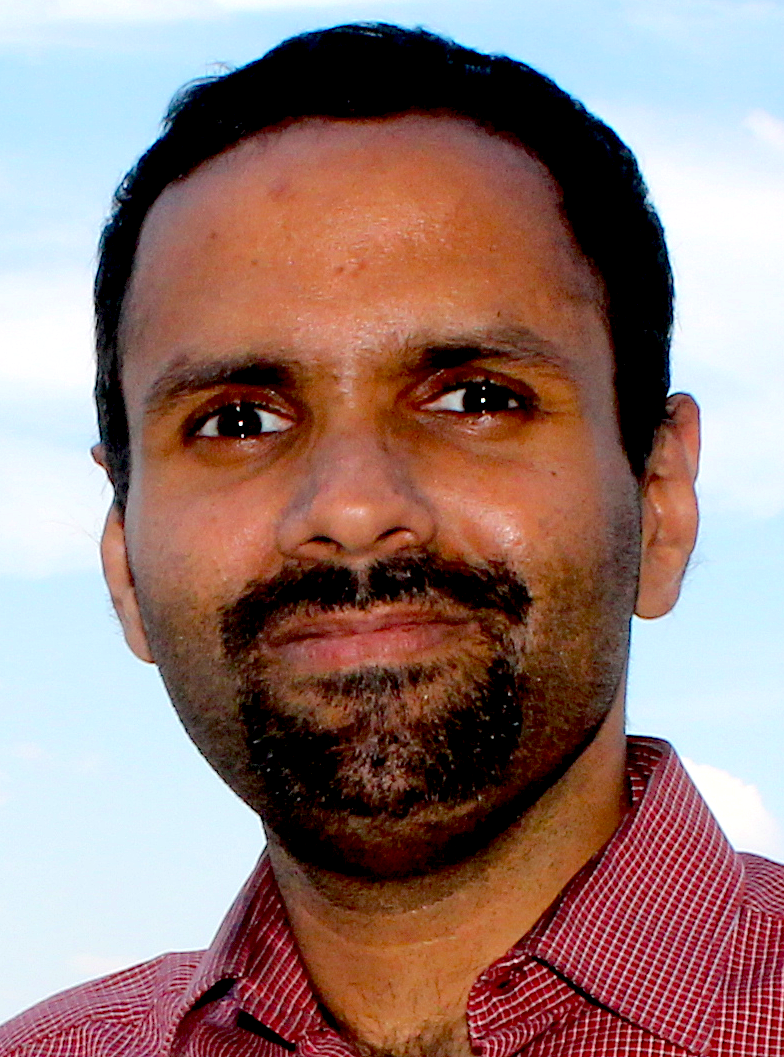}}]{Sriram Ganapathy}
  is a faculty  member  at  the Electrical  Engineering, Indian  Institute  of  Science, Bangalore, where  he  heads  the  activities  of  the learning and extraction of acoustic patterns (LEAP)lab. Prior to joining the Indian Institute of Science,he was a research staff member at the IBM Watson Research Center, Yorktown Heights. He received his Doctor of Philosophy from the Center for Language and  Speech  Processing,  Johns  Hopkins University. He obtained his Bachelor of Technology from College  of  Engineering,  Trivandrum, India and Master of Engineering from  the  Indian  Institute of  Science,  Bangalore.  He has  also worked as a Research Assistant in Idiap Research Institute, Switzerland from 2006 to 2008. At the LEAP lab, his research interests include signal processing, machine learning methodologies for speech and speaker recognition and auditory neuroscience. He is a subject editor for the Speech Communications journal, member of ISCA and senior member of IEEE.
\end{IEEEbiography}







\end{document}